\documentstyle[aps,prl,multicol]{revtex}

\begin{document}
\input epsf
%\draft
\tighten
\preprint{}
\title{
%Dependence of scattering times on  
%the position 
%of th electron distribution along the growth direction in a quantum 
%well\\
%Anomalous density-dependence of scattering times in wide quantum 
%wells\\
Variation of elastic scattering across a quantum well\\
%Wave function displacement: A tool to measure the spatial dependence 
%of elastic scattering
}
\author{G. Salis, P. Wirth, T. Heinzel, T. Ihn, and K. Ensslin\\
}
\address{
Solid State Physics Laboratory, ETH Z\"{u}rich, 8093 Z\"{u}rich, 
Switzerland\\
}
\author{K. Maranowski and 
A. C. Gossard}
\address{
Materials Department, University of California, Santa Barbara, Ca 
93106, USA
\\
}
\date{\today}
\maketitle
\begin{abstract}

The Drude scattering times of electrons in two subbands of a parabolic quantum well 
have been studied at constant electron sheet density and different 
positions of the electron distribution along the growth direction.
The scattering times obtained by magnetotransport
measurements decrease as the electrons are displaced towards the well 
edges, although  the lowest-subband density increases.
By comparing the measurements with calculations of the scattering 
times of a two-subband system, new 
information on the location of the relevant scatterers and the anisotropy 
of intersubband scattering is obtained. It is found that the 
scattering time of electrons in the lower subband depends sensitively on 
the position of the scatterers, which also explains the measured dependence of the 
scattering on the carrier density. The measurements indicate 
segregation of scatterers from the substrate side towards the quantum 
well during growth.
\end{abstract}

\begin{multicols} {2}
\narrowtext
The striking success of Ga[Al]As semiconductor heterostructures 
originates from the extremely high mobilities obtained in these materials. 
One key ingredient for the fabrication of such samples is
modulation doping, where dopants and electrons are spatially separated. 
At low temperatures, impurity scattering, alloy scattering
and interface roughness scattering 
limit the electron mobility \cite{Ando82b}. 
If more than one subband is occupied, intersubband scattering takes place 
in addition\cite{Siggia70,Ando82}. 

Information on the relevant scattering processes is usually obtained 
by measuring how quantum ($\tau_{\rm q}$) and Drude scattering 
times ($\tau$) vary with carrier 
density $n_{\rm S}$. For two-dimensional electron gases (2DEGs) realized in AlGaAs 
heterostructures, it is found that impurity scattering is dominant. 
In this case, one finds $\tau \propto n_{\rm S}^{\gamma}$, with 
$\gamma$ between 1 and 1.5, depending on the distance between the 
dopants and the 2DEG \cite{Ando82b}.

In a two-subband system with subband densities $n_{1}$ and $n_{2}$, the 
Drude  
scattering times $\tau_{i}$ of subband $i$ are usually found to  increase monotonically 
with $n_{i}$ \cite{Smith88b,Zaremba92}. Recent results show that in a 
parabolic quantum well (PQW),
$\tau_1$ may also slowly decrease, i.e. $\gamma<0$, when a second subband is 
occupied \cite{Heinzel98}.
In this paper, we investigate this unusual dependence and show that it may 
be due to a certain arrangement of the ionized impurities.

% sample layout

\begin{figure}
\centerline{\epsfxsize=2.7in \epsfbox{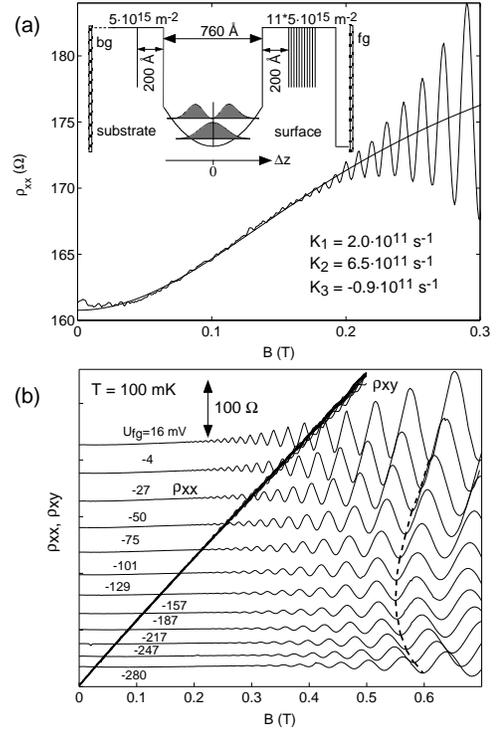}}
\caption{(a) Fit of $\rho_{xx}(B)$ for $V_{fg}$=-50\,mV 
($V_{\rm bg}$=+1000\,mV) to the two-subband scattering model. 
Inset: scheme of sample layout along the 
growth direction. 
(b) Measured $\rho_{xx}$ 
for different electron positions along the growth direction
at $n_{\rm S}=2.9\cdot 10^{15}$\,m$^{-2}$. 
Values for $U_{\rm fg}$  are indicated, and $U_{\rm bg}$ is varied between
-2.2\,V (top) and +2.2\,V (bottom) in steps of 0.4\,V. Subsequent 
data are offset for 
clarity by 50\,$\Omega$. From top to bottom, the electron distribution is displaced 
towards the substrate.  The data for $\rho_{xy}$ fall on top of each 
other since $n_{\rm S}$ is constant. Minima corresponding to the same 
filling factor in the lower subband are connected by a dashed line. }
\label{Fig1}
\end{figure}

The PQW, grown by molecular beam epitaxy (MBE), 
is a 760\,\AA \ wide 
Al$_x$Ga$_{1-x}$As layer with $x$ varying parabolically between 0 and 
0.1 \cite{digital} (inset of Fig.~\ref{Fig1}a). In the center of the well, a three monolayer thick 
Al$_{0.05}$Ga$_{0.95}$As layer forms a potential spike. The well  
is embedded symmetrically in 200\,\AA \ of undoped Al$_{0.3}$Ga$_{0.7}$As spacer 
layers and remote Si-doping layers on both sides.  On the surface side, the donors 
are provided by 11 sheets, each with a Si donor density of 
nominally 5$\cdot$10$^{15}$\,m$^{-2}$ 

% Figure 1

Si-concentration, arranged in a 200\,\AA \ thick layer. On the 
substrate side, the donors are located within one $\delta$-doping layer with a 
concentration of 5$\cdot$10$^{15}$\,m$^{-2}$. 
This asymmetry in the 
doping allows for saturation of the surface states and an effectively 
symmetric location of the electron distribution in the well.
A back gate electrode consists of a 250\,\AA \ thick 
$n^{+}$-doped layer located 1.35\,$\mu$m below the well.  A 
TiPtAu front gate electrode was evaporated on top of the structure. The 
experiments were 
carried out with standard Hall-bar geometries at temperatures of 
100\,mK. A magnetic 
field $B$ was applied perpendicular to the electron gas.

% experiment

Figure~\ref{Fig1}a shows a measurement of the magnetoresistivity 
$\rho_{xx}(B)$ at $n_{\rm S}=2.9\cdot 
10^{15}$\,m$^{-2}$. From the low-field magnetoresistivity, 
$\tau_{1}$ and $\tau_{2}$ are obtained. 
Early studies on scattering times in two-subband systems relied on
the assumption of two independent electronic systems with additive 
conductivities $\sigma=\sigma_{1}+\sigma_{2}$ with $\sigma_{i}=n_{i}e^2\tau_{i}/m$, 
quantitatively explaining a measured positive 
magnetoresistance\cite{Smith88b,Houten88,Smith88} ($e$,$m$ electron 
charge and effective mass). In a more sophisticated  
model based on the Boltzmann equation\cite{Zaremba92}, intersubband scattering is taken 
explicitly into account. This leads to $B$-dependent scattering times

\begin{equation}
\tau_{i}(B)={\rm 
Re}\left(\sum_{j}({\bf{K}}+i\omega_{c}{\bf{1}})^{-1}_{ij}k_{j}/k_{i}\right),
\label{tau1}
\end{equation}

where the $k_{i}$ are the Fermi wave vectors, $k_{i}=\sqrt{2\pi 
n_{i}}$, $\omega_{c}=eB/m$, and $\bf{K}$ the scattering matrix defined by

\begin{equation}
\left(
\begin{array}{cd}
	K1 & K3  \\
	K3 & K2
\end{array}
\right)
=\left(
\begin{array}{cd}
	P_{00}^{(0)}-P_{00}^{(1)}+P_{10}^{(0)} & -P_{10}^{(1)} \\
	-P_{10}^{(1)}& P_{11}^{(0)}-P_{11}^{(1)}+P_{10}^{(0)} 
\end{array}
\right)
\end{equation}

%\begin{eqnarray}
%\bf{K}&=&\left(
%\begin{array}{cd}
%	K1 & K3  \\
%	K3 & K2
%\end{array}
%\right)\\ \nonumber
%&=&
%\left(
%\begin{array}{cd}
%	P_{00}^{(0)}-P_{00}^{(1)}+P_{10}^{(0)} & -P_{10}^{(1)} \\
%	-P_{10}^{(1)}& P_{11}^{(0)}-P_{11}^{(1)}+P_{10}^{(0)} 
%\end{array}
%\right)
%\end{eqnarray} 

The coefficients $P_{nm}^{(i)}$ are related to the transition rates 
$P_{nm}(\phi)$ between subband states $n$ and  $m$ and scattering 
angle $\phi$ by Fourier 
transformation in $\phi$. $P_{ij}^{(0)}$ is the 
transition rate integrated over 
the allowed scattering vectors, while in $P_{ij}^{(1)}$ the integrand is multiplied by 
$\cos \phi$. 
The difference $P_{ii}^{(0)}-P_{ii}^{(1)}$ corresponds to 
the single-subband Drude scattering rate, where the matrix element of the scattering potential 
is weighted by $(1-\cos \theta)$. Note that in the diagonal elements, 
also the isotropic part 
of inter-subband 
scattering is included.
We have shown in a recent paper that intersubband scattering 
cannot be neglected in our experiments\cite{Heinzel98}. 

With $n_{i}$ known,  Eq.~\ref{tau1} allows a 
fit to $\rho_{xx}(B)$, with $K_1, K_2$ and $K_3$ 
being the fit 
parameters\cite{Zaremba92,Heinzel98} (Fig.~\ref{Fig1}a).

We measured $\rho_{xx}(B)$ at $n_{\rm 
S}=2.9\cdot10^{15}$\,m$^{-2}$ (controlled by the low-field Hall 
voltage) 
and different positions of the electron distribution along the 
growth direction (Fig.~\ref{Fig1}b).
The electrons were displaced by 
applying voltages $U_{\rm fg}$ ($U_{\rm bg}$) between the 
front (back) gate 
electrode and the electron 
gas. 

Clearly visible are variations of both amplitude and period 
of the Shubnikov-de Haas (SdH) oscillations with changing $V_{\rm 
fg}$. The 
amplitudes at a fixed magnetic field decay as the wave functions are displaced towards the 
substrate. This corresponds to a decreasing 
$\tau_{q}$\cite{Ando82,Coleridge89}, an effect not to be discussed in 
this paper (see Ref.\onlinecite{Heinzel98} for evaluated data on $\tau_{q}$).
%The $1/B$-frequency of the  
%SdH-oscillations is 
%proportional to $n_{1}$. The absence of a second SdH-frequency 
%indicates a small $\tau_{q}$ in the upper subband.

\begin{figure}
\centerline{\epsfxsize=3.0 in \epsfbox{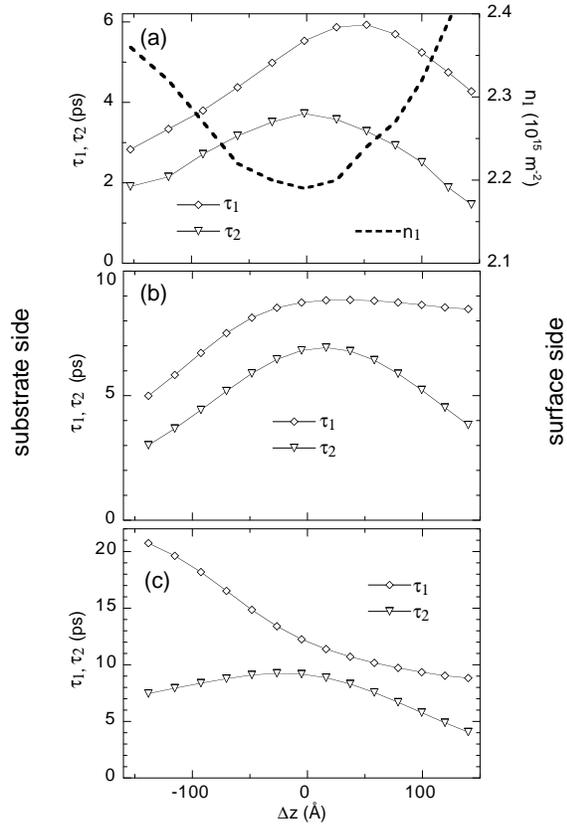}}
\caption{(a) Measurement of $\tau_{1}$, $\tau_{2}$ and $n_{1}$ vs 
$\Delta z$  
at $n_{\rm S}=2.9\cdot 10^{15}$\,m$^{-2}$. 
(b) Calculated scattering times with $1.5\cdot 
10^{15}$m$^{-2}$ of dopants shifted towards the substrate-sided edge 
of the parabolic profile.  
(c) The same calculations as in (b), but with a distribution of 
scatterers as in the growth protocol.}
\label{tau}
\end{figure}

By 
fitting $U_{\rm bg}$ as a function of $U_{\rm fg}$ 
at constant $n_{\rm S}$ to a capacitor model, we find the displacement 
$\Delta z$ per front gate voltage to be 
about 1000\,\AA /V\cite{Salis98b}. Thus we can plot the data as a 
function of $\Delta z$ instead of gate voltages.
From the SdH frequency we evaluate $n_{1}(\Delta z)$ (Fig.~\ref{tau}a). 
A minimum occurs in $n_{1}$ at $U_{\rm 
fg}\approx -130$\,mV and is related 
to the narrow potential spike in the center of the PQW. The spike leads to
subband energy shifts depending sensitively on the electron 
distribution along the growth direction. A displacement of the 
electrons thus changes $n_{1}$ and $n_{2}$. The 
difference between the two lowest subband energies
reaches a minimum when the 
wave functions are centered with respect to the spike. Therefore, the 
minimum in $n_{1}$ provides the reference for the location of the  wave 
functions in growth direction\cite{Salis97}, where $\Delta 
z=0$.

From the data, we evaluated $\tau_{1}$ and $\tau_{2}$ for different $\Delta z$ 
(Fig.~\ref{tau}a). 
Both $\tau_{1}$ and $\tau_{2}$ show a maximum as a function of $\Delta 
z$. The maximum in $\tau_{2}$ occurs  where wave functions are 
centered, i. e. $\Delta z=0$.
 
Assuming a decrease of $\tau_{i}$ with decreasing $n_{i}$ ($\gamma>0$), 
we expect a minimum in 
$\tau_{1}$ at $\Delta z=0$, which disagrees with the measurement.
On the other hand, 
the scattering rate depends on the 
distances from the relevant scatterers \cite{Heiblum84}. For 
$\Delta z=0$, these distances are maximized, giving rise to large $\tau_{i}$. 
The fact that $\tau_{1}$ is large around $\Delta z=0$ indicates that 
not its density-dependence dominates $\tau_{1}$, but 
the distance to the relevant  
scatterers.
In contrast to the first subband, both, $n_{2}$ and $\tau_{2}$ have 
a maximum at $\Delta z=0$. Note that $n_{2}$ is much smaller than $n_{1}$, 
leading to small Fermi wave numbers where screening is more efficient. 
Thus the screened scattering potential at relevant wave 
numbers is
less sensitive to displacements along the growth direction.
On the other hand, the relative change of $n_{2}$ with $\Delta z$ is larger than 
that of $n_{1}$. Hence, $\tau_{2}$ is stronger influenced by its density 
dependence than by $\Delta z$, which explains the coincidence 
of the maximum in $\tau_{2}$ with $\Delta z=0$.

The maximum of $\tau_{1}$ is shifted 
towards the surface, indicating stronger scattering on the substrate side.
Although this could be explained by assuming more dopants than expected from the MBE growth 
protocol, we can
excluded this, because the total amount of Si brought on the wafer was measured accurately. 
However there might be segregation of  dopants on the substrate side
towards the PQW during 
growth, which enhances scattering significantly.
%Another explanation would be a spatial
%correlation of the surface-sided dopants, reducing its scattering 
%efficiency. 
As we 
will show, a calculation of the $\tau_{i}$ supports the assumption of 
segregated Si atoms.

% Figure 2

The matrix elements of the scattering potential were obtained by 
numerical integration using self-consistently calculated wave 
functions\cite{Snider}. Then the transition rates $P_{nm}^{(i)}$ were 
calculated by integrating the 
squared matrix elements over the allowed scattering vectors. Screening 
was included in the Thomas Fermi approximation. The $\tau_{i}$ were 
calculated from Eq.~\ref{tau1}. A detailed calculation of the 
scattering rates based on different scattering mechanisms reveals 
that the contributions of alloy 
scattering (including the potential spike) and interface roughness 
scattering are an order of magnitude smaller than that of Coulomb 
scattering. 
Initially, two layers of Coulomb scatteres were included.  
The dopants on the surface side were gathered in a single $\delta$-layer 300\, \AA\ 
above the well, with a concentration of $N_{1}=3\cdot 10^{16}$\,m$^{-2}$. 
The second layer
is the doping layer 200\,\AA\  below the well ($N_{2}=2.8\cdot 
10^{15}$\,m$^{-2}$).
These values correspond to half 
of the nominal Si concentration  brought on the wafer during the 
MBE-growth, qualitatively accounting for deep donors and not ionized impurities. 
Figure~\ref{tau}b shows the obtained scattering times. 
As expected for this donor configuration, $\tau_{1}$ monotonically 
increases as the electrons 
are displaced towards the substrate side.

In order to take segregated Si atoms into account, we placed 
$N_{3}=1.5\cdot 
10^{15}$\,m$^{-2}$ scatterers at the 
edge 
of the well on the substrate side, 
and reduced $N_{2}$ by the same amount (Fig~\ref{tau}c).
As in the experiment, we obtain a maximum in $\tau_{1}$ displaced towards the surface 
side and a maximum of $\tau_{2}$ at $\Delta z=0$. At the surface 
side, $\tau_{1}$ decreases only slowly, saturating at a value 
comparable to the simulation with $N_{3}=0$. 
It is the balance between the monotonically 
decrasing $\tau_{1}$ shown in Fig.~\ref{tau}c, and the range and 
strength of the extra layer, which determines the exact shape 
of $\tau_{1}(\Delta z)$ 

% Figure 3
\begin{figure}
\centerline{\epsfxsize=2.7 in \epsfbox{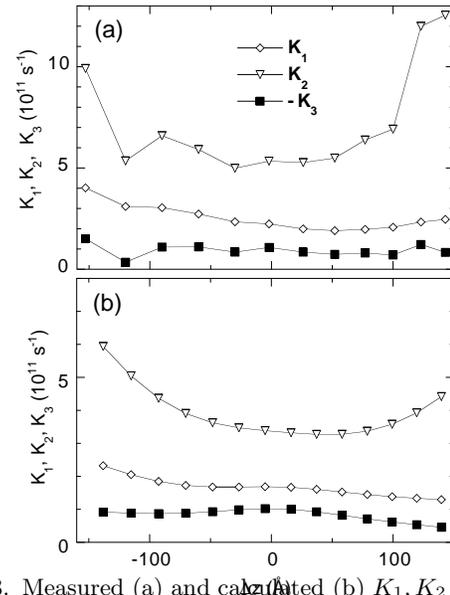}}
\caption{Measured (a) and calculated (b) $K_1, K_2$ 
and $-K_3$. 
In (b), ionized impurity scattering was modelled as in Fig.\ref{tau}c.}
\label{K}
\end{figure}

The calculated scattering times are about 50 percent 
larger than the measured ones. It is well-known that for PQWs 
calculations overestimate the scattering 
times. Possible explanations are size-effect scattering from the edges 
of the electron gas \cite{Walukiewicz91} or enhanced background 
impurities due to the greater reactivity of Al with 
oxygen and carbon-containing molecules in the MBE chamber. In 
addition, the calculated values depend on how screening of the scattering 
potential is implemented and which concentration of ionized impurities 
is assumed.  We did not attempt to simulate  
$\tau_{i}$ accurately, here only the qualitative behavior, in 
particular its spatial dependence, is of 
importance.

Additional insight can be gained by studying the spatial 
variation of the matrix elements $K_{i}$ (Fig.~\ref{K}a). Usually, Drude times 
are insensitive to small-angle scattering. For intersubband 
scattering, $K_{3}$ contains the 
part of the scattering 
rate weighted by $\cosÊ\phi$. This gives information about the amount 
of small-angle intersubband scattering. 
Since almost no structure in $K_{3}$ is observed, while 
$K_{1}$ increases stronger on the 
substrate side, large-angle scattering must be higher on the 
substrate side. In order to 
increase large-angle scattering of Coulomb scatterers with fixed 
density, 
the distance to the electron gas has to be diminished. This
happens if scatterers segregate towards the electron gas.
The
calculated $K_{i}$ nicely reproduce 
the experimental data (Fig~\ref{K}b).

% Figure 4
\begin{figure}
\centerline{\epsfxsize=3.3 in \epsfbox{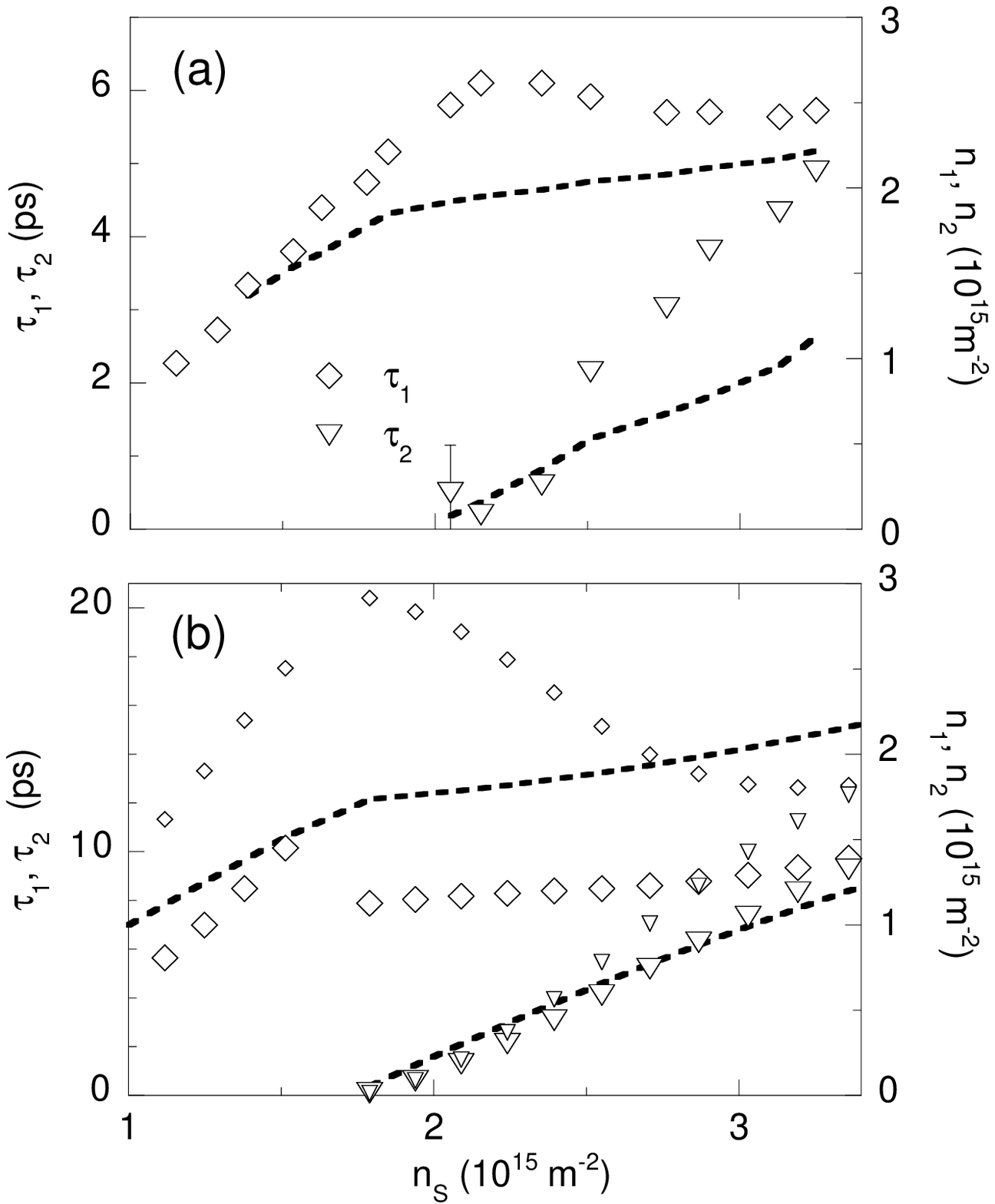}}
\caption{Measurements (a) and calculation (b) of scattering times 
(symbols) and subband 
densities (lines) vs $n_{\rm S}$. In (b), small symbols are 
calculated without, large symbols with additional impurities 
at the substrate side of the well.}
\label{density}
\end{figure}

With this strong evidence for segregated scatterers 
at the substrate side of the well, we come back to the previously unexplained 
structure in the density-dependence of $\tau_{i}$\cite{Heinzel98}. In 
this experiment, $U_{\rm bg}$ was kept fixed, while $U_{\rm fg}$ and 
therefore $n_{s}$ was changed. In 
Fig.~\ref{density}, the measured and calculated values for 
$\tau_{1}, \tau_{2}, n_{1}$ and $n_{2}$ are shown. In the 
measurement, $\tau_{1}$ slightly decreases as $n_{2}$ gets 
populated. In the 
calculation, the additional scattering layer gives rise to a weak 
increase of $\tau_{1}$ with $n_{\rm S}$ when the second subband is 
occupied (large symbols), whereas a 
steep decrease results in the case of no additional layer (small symbols). 
Thus the additional scatters are responsible for the slope of 
$\tau_{1}(n_{\rm S})$. Since $n_{\rm S}$ is driven by $V_{\rm 
fg}$, the electron distribution expands towards the surface side with 
increasing $n_{\rm S}$. Thus the scatterers on both sides of the well
compete and determine the shape of $\tau(n_{\rm S})$. As discussed above, 
for small $n_{2}$, $\tau_{2}$ is not so much
sensitive to additional scatterers, which is reflected in similar values
obtained from the two simulations shown in Fig.~\ref{density}b.

In conclusion, we have presented an investigation of 
Drude scattering times in a modulation-doped multi-subband quantum 
well.
Using front- and back 
gate voltages, the position of the electron distribution and the 
subband densities were tuned.  The Drude scattering times 
of individual subbands were measured. It was found that
$\tau_{1}$ is dominated by the distance of the 2DEG to the impurities 
and not by its density dependence. Its behavior could therefore be 
used to locate additional scatterers at the substrate edge of the well, 
which are presumably due to segregation of dopants during growth.
The measured scattering times
could be qualitatively reproduced in a calculation assuming that half of the 
substrate-sided donors 
had diffused to the edge of the well. Using these results, previous 
measurements of the density dependence of $\tau_{1}$ could be explained.
While obtained for 
a PQW, the presented method of investigating the scattering times as 
a function of the electron-gas position might give further 
informations on 
scattererers in other types of samples. 

We acknowledge valuable discussions with J. Blatter, P. Coleridge, 
K. v. Klitzing, P. Petroff and E. Zaremba. 
This project was financially 
supported by the Swiss Science Foundation and AFOSR grant F 49620-94-1-0158.

%***
%***  References
%***

\end{multicols}


\begin{references}
\bibitem{Snider}We used a one-dimensional Poisson-Schr\"{o}dinger 
solver written by G. Snider.
\end{references}
\end{document}